\documentclass[aps,prb,showpacs,twocolumn,floats]{revtex4}
\usepackage{amssymb}

\usepackage{graphicx}
\usepackage{subfigure}
\usepackage{epsfig}
\usepackage{dcolumn}
\usepackage{bm}
\usepackage{amsmath}

\begin{document}

\bibliographystyle{prsty}
\author{Liufei Cai and E. M. Chudnovsky}
\affiliation{Physics Department, Lehman College, City University of
New York \\ 250 Bedford Park Boulevard West, Bronx, New York
10468-1589, U.S.A.}
\date{\today}

\begin{abstract}
We study electromagnetic interaction of a nanomagnet with a weak
superconducting link. Equations that govern coupled dynamics of
the two systems are derived and investigated numerically. We show
that the presence of a small magnet in the proximity of a weak
link may be detected through Shapiro-like steps caused by the
precession of the magnetic moment. Despite very weak magnetic
field generated by the weak link, a time-dependent bias voltage
applied to the link can initiate a non-linear dynamics of the
nanomagnet that leads to the reversal of its magnetic moment. We
also consider quantum problem in which a nanomagnet interacting
with a weak link is treated as a two-state spin system due to
quantum tunneling between spin-up and spin-down states.
\end{abstract}

\pacs{75.75.Jn, 74.50.+r, 75.45.+j, 03.67.Lx}

\title{Interaction of a Nanomagnet with a Weak Superconducting Link}
\maketitle

\section{Introduction}

Josephson junctions and magnets in a close proximity of each other
can be coupled through various mechanisms. Static properties of
superconductor/ferromagnet/superconductor (S/F/S) Josephson
junctions have been intensively studied in the past but not as
much attention has been paid to the coupled dynamics of the
magnetic moment and the tunneling current. The effect of
superconductivity on ferromagnetic resonance in such junctions has
been recently observed by Bell et al \cite{Bell-2008} who
attributed their observation to the proximity effect
\cite{Buzdin-review}. Theory that may be relevant to this
experiment has been worked out by Buzdin who computed the phase
shift in the Josephson junction arising from the Rashba-type
spin-orbit coupling \cite{Buzdin-2008} and studied the coupled
dynamics of the magnetization and the Josephson current due to
this mechanism \cite{Buzdin-2009}. Dynamical proximity effect
generated by the precession of the magnetization in an S/F/S
junction has been also investigated by Houzet \cite{Houzet-2008}
Shapiro steps in the I-V curve of the S/F/S junction, related to
the ferromagnetic resonance, have been reported by Petkovi\'{c} et
al \cite{Petkovic-2009} who also provided theoretical arguments
favoring a purely electrodynamic nature of the effect in their
experiment.

Coupling of Josephson junctions to individual spins inside the
junction have been also intensively studied in the past. The
theory traces back to the works of Kulik \cite{Kulik-1966} and
Bulaevskii et al \cite{Bulaevskii-1977} who elucidated the effect
of spin flips on the tunneling current. More recently, Nussinov et
al \cite{Zhu-2004,Nussinov-2005}, using Keldysh formalism,
demonstrated that superconducting correlations drastically change
dynamics of a spin inside a Josephson junction. Josephson current
through a multilevel quantum dot with spin-orbit coupling has been
studied by Dell'Anna et al \cite{Anna-2007}. Deposition of a
single magnetic molecule in a SQUID loop has been attempted
\cite{Lam-2008} and theoretical treatments of the Josephson
current through such a molecule have been proposed
\cite{Benjamin-2007,Teber-2010}. Samokhvalov
\cite{Samokhvalov-2009} considered formation of vortices in a
Josephson junction by a magnetic dot. Spin-orbit coupling of a
single spin to the Josephson junction has been studied by
Padurariu and Nazarov \cite{Nazarov-2010} in the context of
superconducting spin qubits. Somewhat related to single spins are
also studies of two-level systems inside Josephson junctions
\cite{two-level}.

Early experimental research on coupling of magnetic nanoparticles
to microSQUIDs has been reviewed by Wernsdorfer
\cite{Wernsdorfer-2001} who also reviewed recent progress made due
to the development of nanoSQUIDs \cite{Wernsdorfer-2009}. Thirion
et al \cite{Thirion-2003} demonstrated the possibility of
switching of the magnetization of 20nm Co nanoparticles in a dc
magnetic field by the radio-frequency pulse generated by a
microSQUID. They measured the angular dependence of the switching
field and reproduced the Stoner-Wohlfarth astroid \cite{Lectures}
for a single nanoparticle. Further miniaturization of such systems
has been achieved using carbon nanotubes \cite{Wern-CNT} and
nanolithography assisted by the atomic force microscope
\cite{Faucher-2009}. Such systems utilizing single magnetic
molecules have been proposed as possible ultimate memory units and
as elements of quantum computers \cite{Wern-08}.

\begin{figure}
\includegraphics[width=70mm]{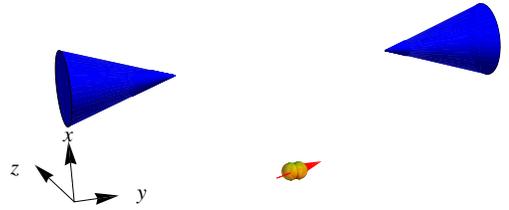}
\caption{Color online: Nanomagnet near a weak superconducting
link.}
\label{junctionmodel}
\end{figure}

In this paper we consider a nanomagnet located close to a weak
link between two superconductors, see Fig.\ \ref{junctionmodel}.
The position of the nanomagnet is away from the path of the
tunneling current, so that the interaction between the two systems
is considered to be of purely electromagnetic origin. The
mechanism of the interaction is conceptually similar to that
argued in the experiment of Ref. \onlinecite{Petkovic-2009}. The
magnetic field of the nanomagnet alters the Josephson current
flowing through the link, while the magnetic flux generated by the
Josephson junction acts on the magnetic moment of the nanomagnet.
From mathematical point of view the dynamics of this problem
resembles the dynamics studied in Ref. \onlinecite{Buzdin-2009}.
The differences stem from different geometry, different
interaction, and finite normal resistance of the weak link that we
allow within the RSJ model. The attractiveness of the problem that
deals with purely electromagnetic interactions is in the absence
of the unknown parameters. We hope that this will assist
experimentalists in designing Josephson junction - nanomagnet
systems with desired properties.

Dynamical equations describing the system depicted in Fig.\
\ref{junctionmodel} are derived in the next Section. Small
oscillations of the magnetic moment of the nanomagnet caused by a
constant voltage applied to the weak link are studied in Section
\ref{Shapiro}. We show that such oscillations can produce
Shapiro-like steps in the I-V curve without any external ac
voltage applied to the link. Non-linear dynamics of the nanomagnet
due to a voltage pulse applied to the link is studied in Section
\ref{Reversal}. We show that using a specific slow time dependence
of the voltage pulse one can reverse the magnetic moment of the
nanomagnet. The remarkable feature of this process is that the
reversal can be achieved despite the fact that the magnetic field
generated by the link is small compared to the switching field
determined by the magnetic anisotropy. This is similar to the
effect of the RF field demonstrated experimentally in Ref.
\onlinecite{Thirion-2003}. The reversal occurs due to the pumping
of spin excitations into the nanomagnet by the ac-field of the
oscillating tunneling current. The final part of the paper studies
electromagnetic interaction of the weak link with a quantum
two-state system formed by tunneling of the nanomagnet's spin
between up and down orientations. Quantum dynamics of this system
is derived in Section \ref{Quantum}. We show that it provides the
simplest realization of a Josephson junction - spin qubit
suggested in Ref. \onlinecite{Nazarov-2010} (see also Ref.
\onlinecite{Wern-08} and Ref. \onlinecite{qubit}). Our conclusions
and suggestions for experiment are summarized in Section
\ref{Discussion}.

\section{The Model}\label{Model}
We consider a system depicted in Fig.\ \ref{junctionmodel}.
Nanomagnet of a fixed-length magnetic moment ${\bf M}$ is located
at a distance $a$ from the center of the weak superconducting link
of length $L$. The nanomagnet is assumed to be rigidly embedded in
the solid matrix of the link. In the presence of the external
magnetic field ${\bf B}_0$, the energy of the nanomagnet is then
given by
\begin{equation}
{\cal{E}}_M = K({\bf M}) - {\bf M}\cdot{\bf B}_0\,,
\end{equation}
where $K({\bf M})$ is the energy of the magnetic anisotropy that
depends on the orientation of ${\bf M}$ with respect to the body
of the magnet.

Neglecting the capacitance of the weak link, the energy of the
link can be written as
\begin{equation}\label{E-J}
{\cal{E}}_J = - E_J\cos \gamma\,,
\end{equation}
where $\gamma$ is the gauge invariant phase, $E_J = \hbar
I_c/(2e)$ is the Josephson energy, and $I_c$ is the critical
current of the link. Note that $E_J$ depends on the external field
${\bf B}_0$. Time derivative of $\gamma$,
\begin{equation}
\frac{d\gamma}{dt} = \frac{2eV(t)}{\hbar}\,,
\end{equation}
is proportional to the total voltage,
\begin{equation}
V(t) = \int_1^2d{\bf r}\cdot {\bf E}({\bf r},t)\,,
\end{equation}
across the link. Here ${\bf E}$ is the electric field and
integration goes from one end to the other end of the link.

For the link biased by the external voltage $V_0(t)$ one has
\begin{equation}\label{gamma}
\gamma = \gamma_0 + \gamma _A\,,
\end{equation}
where
\begin{equation}\label{gamma-0}
\frac{d\gamma_0}{dt} = \frac{2eV_0(t)}{\hbar}
\end{equation}
and
\begin{equation}\label{gamma-A}
\gamma_A = -\frac{2\pi}{\Phi_0}\int_1^2d{\bf r}\cdot{\bf A}({\bf
r},t)\,.
\end{equation}
Here $\Phi_0=2\pi\hbar/(2e)$ is the flux quantum and ${\bf A}$ is
the vector potential.

In our problem the vector potential ${\bf A}$ is formed by two
additive contributions:
\begin{equation}
{\bf A} = {\bf A}_B + {\bf A}_M\,.
\end{equation}
Here
\begin{equation}
{\bf A}_B = \frac{1}{2}({\bf B}\times{\bf r})
\end{equation}
is the vector potential created by the external field and
\begin{equation}\label{A-M}
{\bf A}_M = \frac{\mu_0}{4\pi}\frac{{\bf M}\times{\bf r}}{r^3}
\end{equation}
is the vector potential created at a point ${\bf r}$ from the
nanomagnet assuming that the latter is small compared to all other
dimensions of the problem. The voltage
\begin{equation}
V_A =  \frac{\hbar}{2e}\frac{d\gamma_A}{dt}
\end{equation}
is the electromotive force induced in the link by the
time-dependent magnetic field generated by the rotating magnetic
moment.

The dynamics of the magnetic moment is given by the
Landau-Lifshitz equation:
\begin{align} \label{lleq}
\frac{\partial {\bf M }}{\partial t}  = \gamma_g{\bf M}\times{\bf
B}_{eff}-\frac{\eta}{M_0}|\gamma_g|{\bf M}\times({\bf M
}\times{\bf B}_{eff})
\end{align}
where $\gamma_g$ is the gyromagnetic ratio for ${\bf M}$, $\eta$
is a dimensionless damping coefficient, and
\begin{equation}\label{eff}
{\bf B}_{eff}=-\frac{\partial {\cal{E}}}{\partial{\bf M}}\,
\end{equation}
is the effective field acting on ${\bf M}$, with ${\cal{E}}$ being
the total energy of the system. For ${\cal{E}} = {\cal{E}}_M +
{\cal{E}}_J$ one has
\begin{equation}\label{B-eff}
{\bf B}_{eff} = {\bf B}_0 - \frac{\partial {K}}{\partial{\bf M}} +
I_c \sin\gamma \frac{\partial}{\partial {\bf M}}\int_1^2d{\bf
r}\cdot{\bf A}_M({\bf r},t)\,.
\end{equation}
It is easy to see that the last term in this expression equals the
magnetic field ${\bf B}_J$ created by the tunneling current $I =
I_c \sin\gamma$ at the location of the nanomagnet. Indeed,
substituting into this term ${\bf A}_M$ of Eq.\ (\ref{A-M}) and
rearranging the mixed product of the vectors, one obtains for the
last term in Eq.\ (\ref{B-eff})
\begin{equation}
I_c \sin\gamma\frac{\mu_0}{4\pi}\frac{\partial}{\partial {\bf
M}}\int_1^2 {\bf M}\cdot \frac{{\bf r}\times d{\bf r}}{r^3} =
\frac{\mu_0}{4\pi}\int_1^2\frac{{\bf dI}\times{\bf r}'}{r'^2} =
{\bf B}_J \,,
\end{equation}
where ${\bf r}' = -{\bf r}$ is the radius-vector pointing from the
element of the current to the position of the nanomagnet.

So far we have not considered the normal current through the weak
link. If the resistance of that link, $R$, is finite, the total
current through the link is
\begin{equation}\label{total-I}
I = I_c\sin\gamma + \frac{V}{R}= I_c\sin\gamma + \frac{V_0}{R} +
\frac{\hbar}{2eR}\frac{d\gamma_A}{dt}\,.
\end{equation}
This expression should replace $I_c\sin\gamma$ in the expression
for the effective field, so that in the limit of $I_c \rightarrow
0$ the field given by the last term in Eq.\ (\ref{B-eff}) would be
the field generated by the normal current $I_N = V/R$. Note that
this field can be formally obtained from Eq.\ (\ref{eff}) by
adding the corresponding Zeeman term,
\begin{equation}
{\cal{E}}_Z = - I_N \int_1^2 d{\bf r}\cdot {\bf A}\,,
\end{equation}
to the total energy.

\section{Linear Approximation and Shapiro-like Steps} \label{Shapiro}
In this Section we shall assume that deviations of the magnetic
moment from its equilibrium orientation, caused by the interaction
with the Josephson junction, are small. This will allow us to
treat the Landau-Lfshitz equation in the linear approximation. For
certainty, we choose the external magnetic field ${\bf B}_0$ and
the equilibrium magnetic moment ${\bf M}_0$ in the direction
parallel to the line connecting the leads 1 and 2, which is the
$y$-direction in Fig.\ \ref{junctionmodel}. To make the problem
more tractable we shall also assume in this Section that the
applied field is large compared to the effective field due to
magnetic anisotropy, so that the latter can be neglected.

Under the above assumptions, substitution of Eq.\ (\ref{A-M}) into
Eq.\ (\ref{gamma-A}) gives
\begin{equation}\label{gamma-M}
\gamma_A = -kM_z\,, \qquad k =
\frac{2\pi}{\Phi_0}\frac{L}{a\sqrt{L^2 + a^2}}\,.
\end{equation}
Contribution of the weak link to the effective field is
\begin{equation}
{\bf B}_J = kE_J\left(\sin\gamma + \frac{V_0}{I_cR} - \frac{\hbar
k}{2eI_cR}\frac{dM_z}{dt}\right){\bf e}_z\,,
\end{equation}
where ${\bf e}_z$ is the unit vector in the $z$-direction.
Linearization of Eq.\ ({\ref{lleq}) then gives the following
equation for the perturbation of the magnetic moment in the
$z$-direction:
\begin{eqnarray}\label{m-z}
&&\frac{d^2m_z}{dt^2} + 2\tilde{\eta}\gamma_gB_0\frac{dm_z}{dt} +
\gamma_g^2B_0^2 m_z =\nonumber \\
&&k\gamma_g^2B_0M_0E_J\left(1 + \frac{\eta}{\gamma_gB_0}
\frac{d}{dt}\right) \sin\left(\frac{2eV_0t}{\hbar}-km_z\right)
\end{eqnarray}
where
\begin{equation}
\tilde{\eta} = \eta +
\left(\frac{\hbar}{2e}\right)^2\frac{k^2\gamma_gM_0}{2R}
\end{equation}
is the damping coefficient renormalized by the additional channel
of dissipation due to normal (eddy) currents generated by the
rotating magnetic moment.

According to Eq.\ (\ref{total-I}) the total current is given by
\begin{equation}\label{I-mz}
I = I_c\sin\left(\frac{2eV_0t}{\hbar}-km_z\right) + \frac{V_0}{R}
- \frac{k\hbar}{2eR}\frac{dm_z}{dt}\,.
\end{equation}
It has an ac component and the dc component, $\bar{I}$, that one
can obtain by averaging Eq.\ (\ref{I-mz}) over the oscillations.
We want to compute the dependence of $\bar{I}$ on $V_0$ that is
associated with the I-V curve of the weak link. It is convenient
to introduce
\begin{equation}
\omega_g = \gamma_gB_0\,, \quad \omega_R = \frac{2eI_cR}{\hbar}\,,
\quad E_B = \frac{B_0}{k}\,, \quad \epsilon = \frac{E_J}{E_B}\,,
\end{equation}
and to switch to dimensionless variables:
\begin{eqnarray}
&&\bar{\bf M} = \frac{\bf M}{M_0}\,, \;\; \bar{\bf B}_0 =
\frac{{\bf B}_0}{B_0}\,, \;\; \bar{t} = \omega_g t\,, \nonumber \\
&& \bar{V}_0 = \frac{2eV_0}{\hbar \omega_g}\,, \;\; \bar{I} =
\frac{I}{I_c}\,,  \;\; \bar{k} = kM_0\,.
\end{eqnarray}
Note that $\omega_g$ is the precession frequency for the magnetic
moment in the absence of interaction with the superconducting
link.
\begin{figure}
\includegraphics[width=70mm]{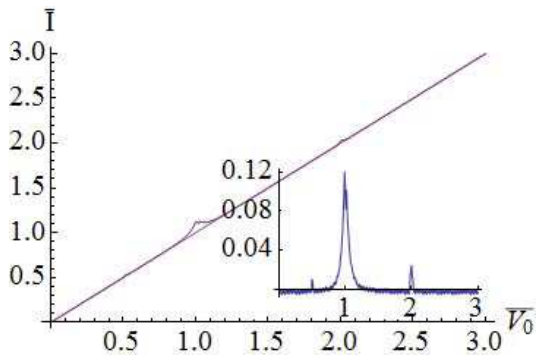}
\caption{Color online: Dependence of $\bar{I}$ on $\bar{V_0}$ in
the presence of the nanomagnet (purple) and without the
magnet(blue) for $\epsilon=0.1$, $\bar{k}=0.1$, $\bar{\eta}
\approx \eta = 10^{-4}$, $\omega_g/\omega_R \ll 1$. The inset
shows the difference between the two curves.}
\label{iv}
\end{figure}

In terms of the above variables the equations for $I$ and $m_z$
become
\begin{equation}\label{Ibar}
\bar{I} = \sin(\bar{V}_0\bar{t} -\bar{k}\bar{m}_z) +
\frac{\omega_g}{\omega_R}\left(\bar{V}_0
-\bar{k}\frac{d\bar{m}_z}{d\bar{t}}\right)\,,
\end{equation}
\begin{equation}\label{mbar-z}
\frac{d^2\bar{m}_z}{dt^2} + 2\tilde{\eta}\frac{d\bar{m}_z}{dt} +
m_z = \epsilon
\left(1+\eta\frac{d}{d\bar{t}}\right)\sin(\bar{V}_0\bar{t}
-\bar{k}\bar{m}_z)\,.
\end{equation}
In these equations the dimensionless parameter $\bar{k}$ can be
small or large, depending on the size and the location of the
magnet. The ratio ${\omega_g}/{\omega_R}$ can also be small or
large depending on the resistance $R$. The parameter $\epsilon$
roughly equals the ratio of the field created by the critical
current at the location of the nanomagnet and the external field.
In practical situations this ratio will always be small, thus,
justifying the linear approximation for $\bar{m}_z$ away from
resonance, $\bar{V}_0 =1$, and at the resonance for not very small
$\tilde{\eta}$. In the case of a very narrow resonance (very small
$\tilde{\eta}$) one should employ the non-linear approximation
based upon the full Landau-Lifshitz equation.

The dependence of $\langle \bar{I} \rangle$ on $\bar{V}_0$,
computed numerically, is shown in Fig.\ \ref{iv}. Shapiro-like
steps at $\bar{V}_0=1$ and $\bar{V}_0=2$ are apparent. They appear
due to same physics as the conventional Shapiro steps, with the
field of the precessing magnet playing the role of the rf field.
The half-Shapiro step that can be seen at $\bar{V}_0=0.5$ appears
when one solves the full Landau-Lifshitz equation instead of the
linearized equation. Fig.\ \ref{iv} illustrates the principal
possibility to detect the presence of a small magnet in the
vicinity of the weak link by measuring its I-V curve.\\

\section{Non-linear Dynamics and Magnetization Reversal} \label{Reversal}

In this Section we demonstrate the possibility of a reversal of
the magnetic moment of the nanomagnet by using a specific time
dependence of the bias voltage applied to the weak link. This
problem involves a non-linear dynamics described by the full
Landau-Lifshitz equation. Consider a nanomagnet with uniaxial
magnetic anisotropy
\begin{equation}
K({\bf M}) = -\frac{K}{2V}M_y^2
\end{equation}
(with $K$ being a constant and $V$ being the volume of the magnet)
in a zero external field. The effective field from this term in
the energy is
\begin{equation}
{\bf B}_{eff}^{(A)} = \frac{KM_y}{V}{\bf e}_y\,.
\end{equation}
The problem of coupling of the weak link to small oscillations of
${\bf M}$ around ${\bf M}_0$ directed along the anisotropy axis
becomes identical to the problem studied in the previous section
if one replaces $B_0$ with $KM_0/V$. If both are present $B_0$ in
the above formulas should be replaced with $B_0 + KM_0/V$.

The dc magnetic field that would be required to switch the
magnetic moment to the opposite orientation along the anisotropy
axis is $B_0=KM_0/V $. In all practical situations the magnitude
of the oscillating magnetic field produced by the weak link at the
location of the magnet will be hopelessly small compared to that
field. The question, however, arises whether the ac field produced
by the oscillations of the Josephson current can pump spin
excitations into the magnet at a rate sufficient to reverse its
magnetization. We shall see that this may, indeed, be practicable,
thus invoking the possibility of a magnetic memory unit operated
by voltage pulses.

When the effective field is dominated by the magnetic anisotropy
the roles of the parameters $\omega_g$ and $E_B$ are played by
\begin{equation}
\omega_g = \frac{{\gamma}_gKM_0}{V}\,, \quad E_B =
\frac{KM_0}{kV}\,.
\end{equation}
To simplify our formulas we consider in this Section the limit of
a very large normal resistance, so that we can neglect the normal
current through the link. (This assumption is unessential for our
conclusions, though, and the calculation can easily be generalized
to the case when the normal current is present.) Under this
assumption the non-linear dynamics of the magnetic moment is
described by the dimensionless Landau-Lifshitz equation
\begin{align} \label{lleq-dim}
\frac{\partial \bar{\bf M }}{\partial \bar{t}}  = \bar{\bf
M}\times\bar{\bf B}_{eff}-{\eta}\bar{\bf M}\times(\bar{\bf M
}\times\bar{\bf B}_{eff})
\end{align}
with dimensionless
\begin{equation}
\bar{\bf B}_{eff} = \bar{M}_y{\bf e}_y +
\epsilon\sin(\bar{V}_0\bar{t} -\bar{k}\bar{M}_z){\bf e}_z\,.
\end{equation}

Numerical solution of Eq.\ (\ref{lleq-dim}) for a time-linear
voltage pulse is shown in Fig.\ \ref{reversal1}.
\begin{figure}
\includegraphics[width=70mm]{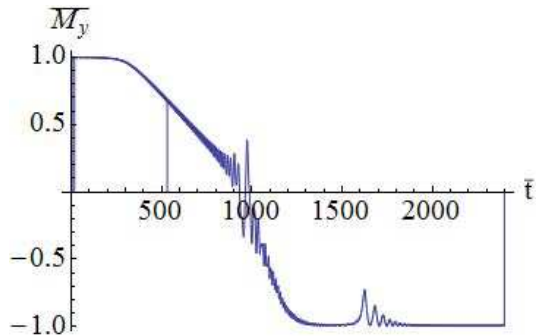}
\caption{Color online: Reversal of the magnetic moment of the
nanomagnet by linearly decreasing bias voltage
$\bar{V}=1.5-0.00075\bar{t}$. The values of the parameters are
$\epsilon = \bar{k} = 0.05$, $\eta = 0.01$.} \label{reversal1}
\end{figure}
As the effective field decreases in the course of the reversal,
the ac field generated by the oscillating tunneling current
continuously pumps spin excitations into the nanomagnet. This
leads to the full reversal of the magnetic moment. We find
numerically that the reversal only occurs at $\epsilon > \eta$.
Another observation is that the time needed for the reversal is
inversely proportional to $\epsilon$. Practical implications of
these findings are discussed in Sec.\ \ref{Discussion}.

\section{Nanomagnet as a Two-State Quantum System} \label{Quantum}

In this Section we will treat nanomagnet as a fixed-length quantum
spin ${\bf S}$, rigidly embedded in a solid matrix. Magnetic
anisotropy energy $K({\bf M})$ should now be replaced by a
crystal-field Hamiltonian. The general form of such a Hamiltonian
that corresponds to a strong easy-axis magnetic anisotropy is
\begin{equation}\label{ham}
\hat{H}_S = \hat{H}_{\parallel} + \hat{H}_{\perp}\,,
\end{equation}
where $\hat{H}_{\parallel}$ commutes with $S_z$ and
$\hat{H}_{\perp}$ is a perturbation that does not commute with
$S_z$. Presence of the magnetic anisotropy axis means that the $|
\pm S\rangle$ eigenstates of $S_z$ are degenerate ground states of
$\hat{H}_{\parallel}$. Operator $\hat{H}_{\perp}$ slightly
perturbs the $| \pm S\rangle$ states, adding to them small
contributions of other $|m_S\rangle$ states. We shall call these
degenerate normalized perturbed states $|\psi_{\pm S}\rangle$.
Physically they describe the magnetic moment of the nanomagnet
looking in one of the two directions along the anisotropy axis.
Full perturbation theory with account of the degeneracy of
$\hat{H}_{S}$ provides quantum tunneling between the $|\psi_{\pm
S}\rangle$ states \cite{MQT-book}. The ground state and the first
excited state are even and odd combinations of $|\psi_{\pm
S}\rangle$ respectively,
\begin{equation}\label{pm}
\Psi_{\mp} = \frac{1}{\sqrt{2}}\left(|\psi_{S}\rangle \pm
|\psi_{-S}\rangle\right)\,.
\end{equation}
They satisfy
\begin{equation}\label{Epm}
\hat{H}_S\Psi_{\pm} = E_{\pm}\Psi_{\pm}
\end{equation}
with
\begin{equation}
E_+ - E_- = \Delta
\end{equation}
being the tunnel splitting. The latter is typically small compared
to the distance to other spin energy levels, making the two-state
approximation rather accurate at low energies. For, e.g., biaxial
magnetic anisotropy, $\hat{H}_S = -DS_z^2 + dS_y^2$ with $d \ll
D$, the splitting of the lowest energy level appears in the
$S$-order on $d/D$, while the distance to the next level equals
$(2S-1)D$.

\begin{figure*}
\subfigure[$ \; \hbar\omega_J=\Delta$]{
\includegraphics[width=70mm]{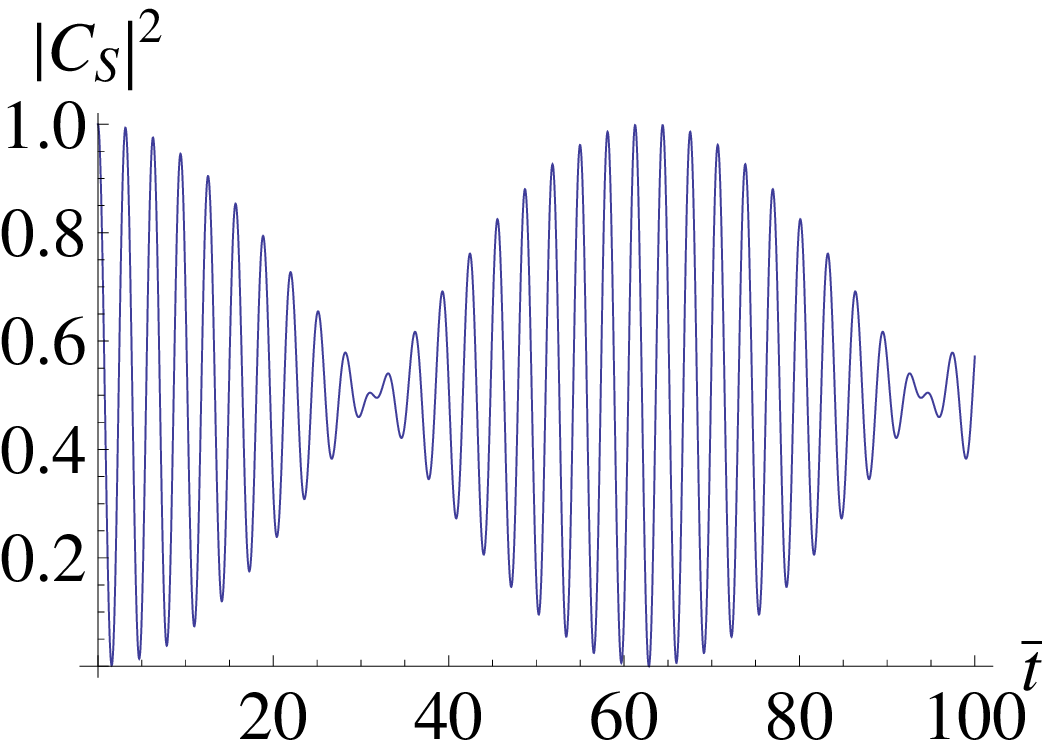}
\label{qubit} } \subfigure[$\;\hbar\omega_J=3\Delta/2$]{
\includegraphics[width=70mm]{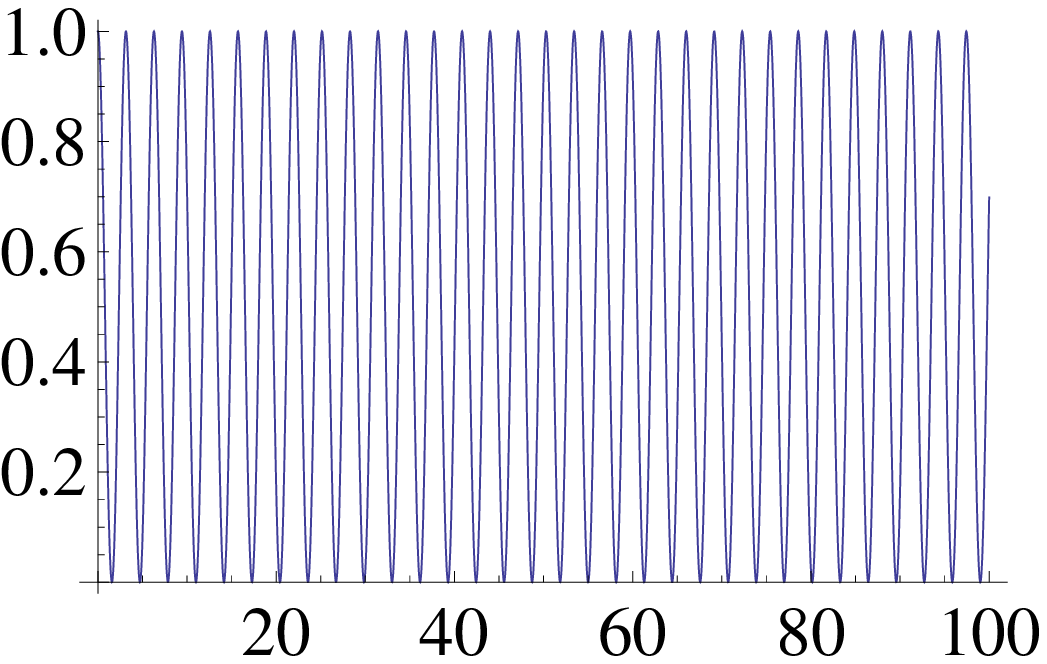}
} \caption{Color online: Rabi oscillations of $|{C}_S(\tau)|^2$ at
${2\bar{k}E_J}/{\Delta}=0.1$ for two constant voltages with the
initial condition ${C}_S(0)=1$.} \label{qbts}
\end{figure*}
Since the two low-energy spin states of quantum nanomagnet are
superpositions of $|\psi_{\pm S}\rangle$, it is convenient to
describe such a two-state system by a pseudospin 1/2. Components
of the corresponding Pauli operator ${\bm \sigma}$ are
\begin{eqnarray}\label{Pauli}
\sigma_x & = & |\psi_{-S}\rangle\langle \psi_{S}| +
|\psi_{S}\rangle\langle \psi_{-S}|
 \nonumber \\
 \sigma_y & = &
i|\psi_{-S}\rangle\langle \psi_{S}| - i|\psi_{S}\rangle\langle
\psi_{-S}| \nonumber \\ \sigma_z & = & |\psi_{S}\rangle\langle
\psi_{S}| - |\psi_{-S}\rangle\langle \psi_{-S}|\,.
\end{eqnarray}
The projection of any operator $\hat{A}_S = \hat{A}(\hat{\bf S})$
onto $|\psi_{\pm S}\rangle$ states is
\begin{equation}\label{projection}
\hat{A}_{\sigma} = \sum_{m,n = \psi_{\pm S}}\langle
m|\hat{A}_S|n\rangle|m\rangle\langle n|\,.
\end{equation}
Expressing $|\psi_{\pm S}\rangle$ via $\Psi_{\pm}$ according to
Eq.\ (\ref{pm}), it is easy to see from Eq.\ (\ref{Epm}) that
\begin{equation}\label{ME}
\langle \psi_{\pm S}|\hat{H}_S|\psi_{\pm S}\rangle =  0, \quad
\langle \psi_{-S}| \hat{H}_S|\psi_{S}\rangle = - {\Delta}/{2}\,.
\end{equation}
With the help of these relations one obtains from Eq.\
(\ref{projection})
\begin{equation}\label{projection1}
\hat{H}_{\sigma}^{(S)} = -({\Delta}/{2})\sigma_x\,.
\end{equation}

Quantum generalization of Eq.\ (\ref{E-J}) with account of Eqs.\
(\ref{gamma-0}) and (\ref{gamma-M}) is
\begin{equation}
\hat{H}_J = -E_J\cos(\omega_J t - k\mu_B \hat{S}_z)\,.
\end{equation}
Equations (\ref{Pauli}) and (\ref{projection}) then give
\begin{equation}
\hat{H}_{\sigma}^{(J)} = -E_J[\cos(\omega_J t) + k\mu_B S \sigma_z
\sin(\omega_J t)]\,,
\end{equation}
where
\begin{equation}
\omega_J(t) = \frac{2e}{\hbar}V_0(t)\,.
\end{equation}
The total Hamiltonian of our two-level system is
\begin{equation}
\hat{H}_{\sigma} = -{\bf b}_{eff}\cdot {\bm \sigma}\,,
\end{equation}
where
\begin{equation}
\hat{\bf b}_{eff} = ({\Delta}/{2}){\bf e}_x +
\bar{k}E_J\sin(\omega_J t){\bf e}_z\,.
\end{equation}
Here, as before, $\bar{k} = kM_0 = k\mu_B S$, so that the energy
$\bar{k}E_J$ roughly represents the strength of the interaction of
the magnetic flux of the junction with the spin ${\bf S}$. In
practice this interaction can be greater or smaller than $\Delta$.

The above system represents a simple realization of the spin qubit
proposed by Nazarov \cite{Nazarov-2010}. Quantum states of such a
qubit are described by the wave function
 \begin{equation}\label{psi}
\Psi =C_{S}|\psi_{S}\rangle+C_{-S}|\psi_{-S}\rangle\,.
 \end{equation}
 The Schr\"{o}dinger equation for $ |\Psi\rangle$ is
 \begin{equation}\label{Schrodinger}
 i\hbar\frac{d\Psi}{dt} = \hat{H}_{\sigma}\Psi\,.
 \end{equation}
 Here $|C_{\pm S}|^2$ is the probability for the spin to look up
 or down along the $z$-direction, with $|C_{+S}|^2 +
 |C_{-S}|^2=1$.
 Introducing
 \begin{equation}
\tilde{C}_{\pm S} (t)= C_{\pm S
}(t)\exp\left\{\frac{i\bar{k}E_J}{2\hbar}\int_{t_0}^{t}dt'sin[\omega_J(t')
t']\right\}\,,
\end{equation}
we obtain from Eq.\ (\ref{Schrodinger})
 \begin{align}\label{cs}
 i\hbar\frac{d}{dt}\tilde{C}_{-S}&=\frac{\Delta}{2}\tilde{C}_{S}\nonumber\\
 i\hbar\frac{d}{dt}\tilde{C}_{S}=&-\bar{k}E_J\tilde{C}_S\sin(\omega_J t)
 +\frac{\Delta}{2}\tilde{C}_{-S}\,.
 \end{align}
In terms of dimensionless time, $\bar{t}={\Delta t}/({2\hbar})$,
the resulting equation for $\tilde{C}_S$ is
 \begin{equation}\label{cs*}
 \frac{d^2}{d{\bar{t}}^2}\tilde{C}_S-i\frac{2\bar{k}E_J}{\Delta}\frac{d}{d\bar{t}}
 \left[\sin\left(\frac{2\hbar\omega_J}{\Delta}\bar{t}\right)\tilde{C}_S\right]+\tilde{C}_S=0
 \end{equation}
The effect of the bias voltage becomes especially pronounced at
$V_{0}$ satisfying $\omega_J = (m/n)\Delta$, where $m$ and $n$ are
integers. Fig.\ \ref{qbts} shows Rabi oscillations of the
probability to remain in the initial spin-up state for two
different bias voltages each satisfying one of the above
conditions.
\\

\section{Discussion} \label{Discussion}

We have studied electromagnetic interaction between a weak
superconducting link and a small magnet placed in the vicinity of
the link. Three problems have been considered: Shapiro-like steps
in the I-V curve generated by the magnet, the reversal of the
magnetic moment by a time-dependent bias voltage, and Rabi
oscillations of the quantum spin induced by a constant voltage.

For the first two problems the strength of the interaction is
determined by the parameter $\epsilon={E_J}/{E_B}$. By order of
magnitude ${\epsilon}$ represents the ratio of the magnetic field
generated by the tunneling current and the effective field,
$B_{eff}$, acting on the magnetic moment due to magnetic
anisotropy and the applied external field. We show that precession
of the magnetic moment generates Shapiro-like steps in the I-V
curve of the superconducting weak link. The possibility to observe
the first Shapiro step at $\bar{V}_0=1$ (and also the peak at $V_0
= 0.5$ due to non-linearity) appears quite realistic. Note that
the first step scales down linearly with $\epsilon$ when
decreasing $\epsilon$, the second step at $\bar{V}_0=2$ scales as
$\epsilon^2$, and so on. Thus, for small $\epsilon$, higher steps
may be more difficult to see in experiment.

A remarkable observation is that despite the weakness of the field
generated by the tunneling current of the link, for a certain time
dependence of the bias voltage it can effectively pump spin
excitations into the magnet, leading to the reversal of its
magnetic moment. The damping constant $\eta = 0.01$ was chosen for
simulations of the reversal. This value is realistic for magnetic
nanoparticles \cite{Coffey-1998}. We find that condition $\epsilon
> \eta$ is required for the reversal, which must be important for
experiment. The parameter $\epsilon$ determines the number of
cycles in the precession of the magnetic moment that leads to the
reversal of the moment. In our numerical simulations that number
roughly scaled as $1/\epsilon$. For $\epsilon = 0.05$ used to
obtain the plot shown in Fig.\ \ref{reversal1} the time required
to reverse the moment was close to $10^{3}\omega_g^{-1}$. For,
e.g., $\omega_g \sim 10^{11}$s$^{-1}$ this would provide a
reversal in ten nanoseconds. The linear time dependence of the
bias voltage in Fig.\ \ref{reversal1} was chosen to maintain the
condition of continuous pumping of spin excitations into the
magnet. Smaller $\epsilon$ would require slower time dependence of
$V_0$, which should not be difficult to satisfy in experiment.
However, smaller $\epsilon$ would require smaller $\eta$ due to
the condition $\epsilon > \eta$. Also, the smaller is $\epsilon$
the more sensitive the time evolution of the magnetic moment
becomes to the time dependence of the voltage. A slight change in
that dependence is sufficient for the moment to bounce back to the
initial direction after reaching the top of the anisotropy
barrier.

In the quantum problem, the parameter $\epsilon$ is no longer
relevant. The relevant parameter becomes the ratio of the Zeeman
interaction of the spin with the field of the tunneling current
and the tunnel splitting $\Delta$. This parameter can be small or
large depending on the splitting. Rabi oscillations of the spin
are strongly affected by the bias voltage. The most noticeable
effect appears at $V_0$ satisfying one of the resonant conditions
$eV_0 = (m/n)\Delta$, where $m$ and $n$ are integers. At such
resonances the behavior of the probability to find the spin in up
or down configurations is very different from the off-resonance
behavior. This demonstrates the principal possibility to
electromagnetically manipulate a nanomagnet - weak link qubit by
the voltage applied to the link.

\section{Acknowledgements}
The authors acknowledge invaluable discussions with Wolfgang
Wernsdorfer. This work has been supported by the U.S. Department
of Energy through Grant No. DE-FG02-93ER45487.

\end{document}